\documentclass[aps,prd,twoside,twocolumn,floatfix,letterpaper]{revtex4-1}
\usepackage[portuguese]{babel}
\usepackage[latin1]{inputenc}
\usepackage[dvipdfm]{graphicx}
\usepackage[dvipdfm]{rotating}
\usepackage{graphicx,pdfpages}
\usepackage{subfig}
\usepackage{float}
\usepackage{lipsum,fancyhdr}
\usepackage{textcomp}
\usepackage{subfig}
\usepackage{array}
\usepackage{amsmath,amssymb}
\usepackage[T1]{fontenc}
\usepackage{mathptmx}

\fancypagestyle{plain}{%
\chead{}

}

\newcommand{\hcond}{\usefont{T1}{phv}{mc}{n}} 
\makeatletter
\def\section{\@startsection{section}{1}{0pt}{-3.5ex plus -1ex minus -.2ex}{2.3ex plus .2ex}{\large \hcond}}
\def\subsection{\@startsection{subsection}{2}{\z@}{-3.25ex plus -1ex minus -.2ex}{1.5ex plus .2ex}{\hcond}}

\begin{document}

\setcounter{page}{1}

\title{Identifying Patterns on Cosmic Ray Maps with Wavelets on the Sphere}
\author{Rafael Alves Batista}\email[]{rab@ifi.unicamp.br}
\author{Marcelo Zimbres}
\author{Ernesto Kemp}
\affiliation{Instituto de F\'isica ``Gleb Wataghin'' - Universidade Estadual de Campinas}


\begin{abstract}
The deflection of ultra-high energy cosmic rays depends on the shape of the injection spectrum of the source and the pervasive cosmic magnetic fields. In this work it is applied the wavelet transform on the sphere to search for energy ordered filamentary structures arisen from magnetic bending. These structures, the so-called multiplets, can bring relevant information concerning the intervening magnetic fields.
\end{abstract}

\maketitle\thispagestyle{plain}

\section{Introduction}

Ultra-High Energy Cosmic Rays (UHECRs) are particles with energy above 10$^{18}$ eV reaching the Earth. Almost a century after the discovery of cosmic radiation, the origin, chemical composition and mechanisms of acceleration and propagation of the UHECRs remain a mystery\cite{nagano00}.

The way their arrival directions are represented is by associating each event with its corresponding latitude and longitude in the celestial sphere. For charged particles, the pervasive cosmic magnetic fields play an important role, affecting the distribution of arrival directions of UHECRs.

Several mechanisms of acceleration of UHECR in different astrophysical sites result on a power law differential energy spectrum\cite{protheroe99}, i. e.,
\begin{equation}
	\frac{dN}{dE} \propto E^{-\alpha},
\end{equation}
where $\alpha$ is the spectral index. For a single source emitting UHECRs of energy $E$ according to this spectrum, the expected angular deflection is inversely proportional to the energy. Therefore, particles with different energies will deflect differently and the greater the energy of the particle the smaller the angular deflection of the particle (and vice-versa). So, if locally, the typical scale of the intervening magnetic fields (coherence length) is smaller or of the same order of magnitude of the Larmor radius of the cosmic ray, filamentary structures in sky maps are expected to be formed. The cosmic ray events that compose this structure are ordered by energy, creating the so-called multiplets of UHECRs. For the galactic magnetic field the expected angular deflection ($\delta$) is given by\cite{giacinti10}
\begin{equation}
	\delta = 8.1^\circ 40\frac{Z}{E} \left| \int\limits_0^L \frac{d\vec{r}}{3{\ }\mbox{kpc}} \times \frac{\vec{B}}{2{\ }\mbox{kpc}}  \right|,
\end{equation}
where $L$ is the distance to the source, $\vec{B}$ is the magnetic field, $Z$ is the atomic number of the nuclei and $E$ its energy in EeV.

\section{The Method}

\par In this work it is proposed a new method (for another method refer to \cite{auger11}) to identify multiplets in maps containing arrival directions of UHECRs. This method is based on the spherical wavelet transform\cite{wiaux08} and consists on the correlation of a signal with rotated versions of a given pattern, both defined on a spherical manifold. Let $a_{lm}$ be the coefficients of the spherical harmonic expansion of the signal, and $b_{ln}$ be the same expansion for the sought rotated pattern (in this case, a wavelet). The correlation coefficient is
\begin{equation}
	C =\sum\limits_{l=0}^{B-1} \sum\limits_{|m|\leq l} \sum\limits_{|m'|\leq l} \overline{a_{lm}}b_{lm'}D^l_{mm'}(\alpha,\beta,\gamma),
\end{equation}
where the overline indicates the complex conjugate, $D^l_{mm'}$ designates Wigner-D functions, $B$ is the band limit of the implemented algorithm, and $(\alpha,\beta,\gamma)$ are the Euler angles associated to the rotation of the wavelet. 

\par Applying the spherical wavelet transform using a code called SWAT (Spherical Wavelet Analysis Tool)\cite{swat} to the distribution of arrival directions of UHECRs, if the correlation coefficient is high, this indicates a similarity between the pattern (wavelet) used and the signal. The Euler angles provide information regarding the coordinates (latitude and longitude) where there is a maximal correlation coefficient, and the orientation of the pattern.  

\par Even though the method is able to identify filamentary structures, it does not take into account the energy ordering of the events for the analysis. This is done by choosing a rectangle of dimensions $\sim$ 12$^\circ$ $\times$ 1.5$^\circ$ (typical dimensions of a multiplet) with the obtained orientation, and calculating the Pearson coefficient for $\delta \times E^{-1}$ of the events within this rectangle. If this coefficient is high, it indicates an energy ordering and the filamentary structure is a possible multiplet candidate.

\section{Application to Simulations}

\par Events of UHECRs were simulated using the code CRT\cite{crt}, which numerically solves Lorentz force equations of motion for charged particles in a magnetic field using fifth order adaptative Runge-Kutta routines. Fifty proton events were simulated from a hypothetical source lying at a distance of 4 kpc from Earth, with galactic coordinates $(l,b)=(90^\circ,-45^\circ)$. The source was assumed to have an injection spectrum with $\alpha=2.7$ and to emit particles within a perfectly collimated jet. 

\par The magnetic field models used in the simulation are the ones proposed by Harari, Mollerach and Roulet\cite{harari99}, namely the ASS-S, ASS-A, BSS-S and BSS-A models, were ASS stands for AxiSymmetric Spiral (even under the transformation $\theta \rightarrow \theta +\pi$, where $\theta$ is the azimuth angle) and BSS for BiSymmetric Spiral (odd with respect to the mentioned transformation). The sufixes -A and -S indicate the parity of the transformation above and below the galactic disk, where -S is symmetric (even) and -A is antisymmetric (odd) with respect to the $z\rightarrow z$ transformation. In this class of models the magnetic field in the disk of the galaxy can be decomposed in a radial and an azimuthal part, depending on the pitch angle ($p$) of the logarithmic spiral that models the magnetic field of the Milky Way. The intensity of the magnetic field in the disk is
\begin{equation}
	B(r,\theta) = 3\frac{r_0}{r} \tanh^3\left(\frac{r}{r_1}\right) \cos^m\left(  \theta - \frac{1}{\tan p} \ln \frac{r}{r_0} \right){\  } \mu G,
\end{equation}
where $r_0=10.55{\ }\mathrm{kpc}$, $r_1=2{\ }\mathrm{kpc}$, $p=-10^\circ$ and $m$ is 1 for BSS models and 2 for ASS models. The $z$ component of the magnetic field is 
\begin{equation}
	B(z)=\left[ 2\cosh\left( \frac{z}{z_0}\right) + 2\cosh\left( \frac{z}{z_1}\right)\right]^{-1} \zeta(z) {\   }{\  } \mu G,
\end{equation}
where $\zeta(z)=\tanh(z/z_2)$ for -A models and $1$ for -S models, $z_0=4{\  }\mathrm{kpc}$, $z_1=0.3{\  }\mathrm{kpc}$ and $z_2=20{\  }\mathrm{pc}$.

\par
\begin{figure}[h!]
	\includegraphics[scale=0.45]{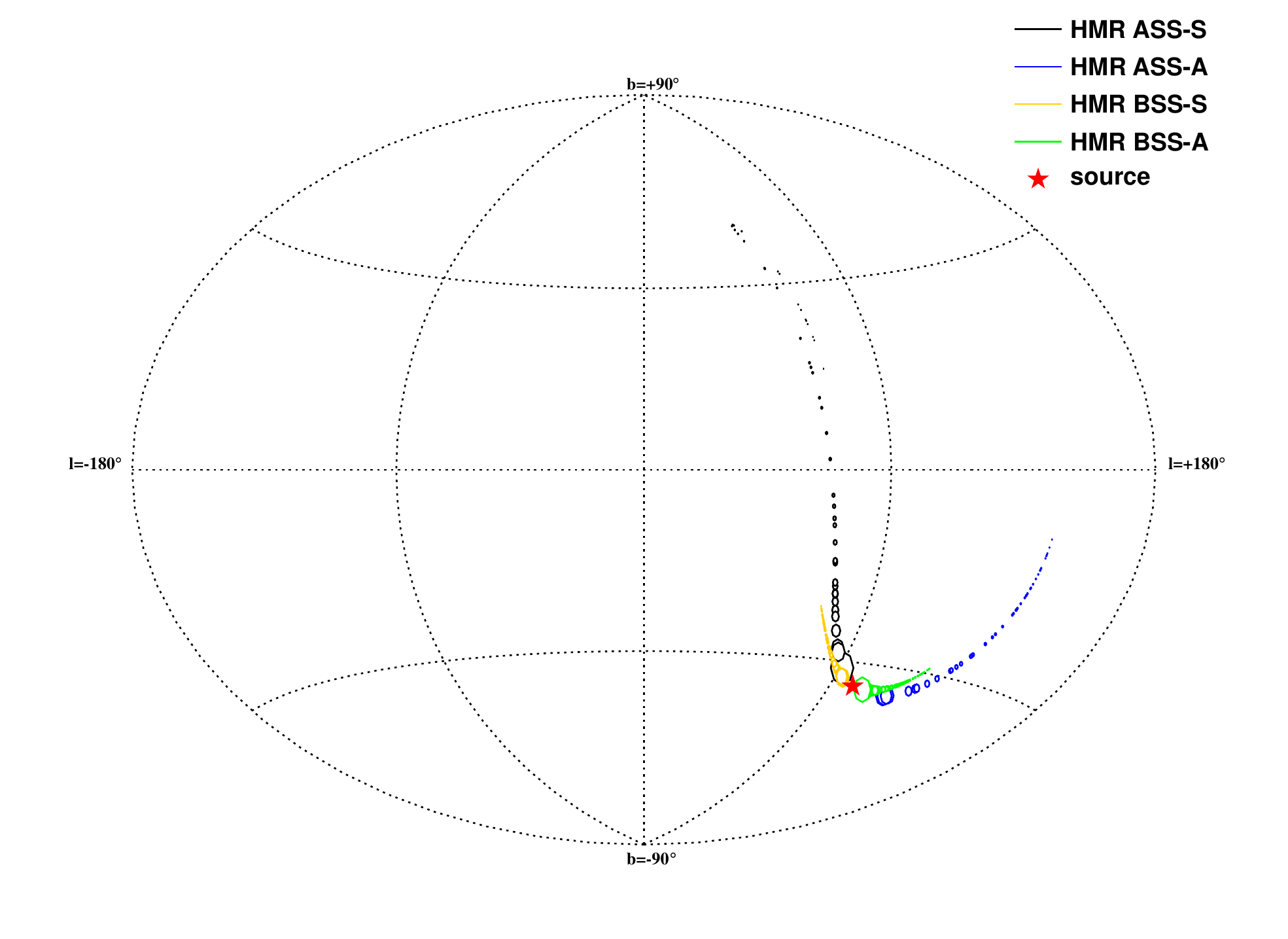}
	\caption{Skymap containing the arrival direction of the simulated events. The size of the circles are proportional to the energy of the event.}
	\label{fig:mult}
\end{figure}

 By applying the method to these specific simulations, using the wavelet transform with suitable parameters (compatible with the expected dimensions of a multiplet), the multiplet was correctly identified and the orientation for different models was calculated and is shown in table \ref{tab:tab}. Figure \ref{fig:mult} shows the simulated events for the simulations.

\begin{table}[h!]
	\label{tab:tab}	\caption{Orientation ($\chi$) for the simulated source in the class of models of Harari, Mollerach and Roulet, for the different symmetries.}
	\begin{tabular}{cc}
	Model & $\chi$ ($^\circ$) \\
	\hline
	ASS-S & 100.3 \\
	ASS-A & 108.4 \\
	BSS-S & 65.7 \\
	BSS-A &  68.1 \\
	\end{tabular}
	\label{tab:tab}
\end{table}

\section{Conclusions and Perspectives}

\par The orientation of the multiplet, i.e., the angle it forms with the galactic plane, depends on the coordinates of the source and the magnetic field model adopted. Therefore, it is possible to set limits on models of the galactic magnetic field by analysing the orientation of multiplets. When the method was applied to simulated data, the multiplets were correctly identified and it was noticed a relation between the orientation of the multiplet and the adopted model. If this method is applied to data collected by the Pierre Auger Observatory or any other cosmic ray experiment and is able to successfully identify multiplets, a new tool to probe the galactic magnetic field will be available.

\section*{Acknowledgements}
\vspace{-0.3cm}
We are grateful for the financial support of FAPESP (Funda\c c\~ao de Amparo \`a Pesquisa do Estado de S\~ao Paulo) and CAPES (Coordedoria de Aperfei\c coamento de Pessoal de N\'ivel Superior).



\end{document}